\documentclass[12pt]{article}
\usepackage{color}
\usepackage{latexsym}
\usepackage{amsmath}
\usepackage{amsfonts}
\usepackage{amssymb}
\usepackage{indentfirst} 
 
\newcommand{\mR}{{\mathbb R}}

\title{Nonrelativistic conformal transformations in Lagrangian formalism}
\author{K. Andrzejewski\thanks{e-mail: k-andrzejewski@uni.lodz.pl},
J. Gonera, A. Kijanka-Dec\\
\small Department of Theoretical Physics and Computer Science, \\
\small University of \L\'od\'z,\\
\small Pomorska 149/153, 90-236 {\L}\'od\'z, Poland
}
\date{}
\begin{document}
\maketitle 
\begin{abstract}
The conformal transformations corresponding to $N$-Galilean conformal symmetries, previously defined as canonical symmetry transformations on phase space, are constructed as point transformations in coordinate space.
\end{abstract}
\section{Introduction}
In the recent paper \cite{b1} we have constructed, using the orbit method \cite{b2}, the general Hamiltonian system on which the centrally extended $N$-Galilean conformal algebra \cite{b3} (group) acts transitively  as symmetry group (the central extension is possible for $N$ - odd in any space-time and for all  $N$  in $(2+1)$-dimensional one \cite{b4}-\cite{b6}). It appears that any such system consists of "external"  variables forming standard phase space and the "internal" ones: spin (related to $SU(2)$  subgroup) and pseudospin (related to $SL(2,\mR)$  subgroup). The dynamics of external and internal variables are decoupled. Neglecting the internal variables one obtains the free dynamics governed by higher-derivative  theory {\cite{b6} (in fact, the necessity of considering higher-derivative theory was revealed  in \cite{b7,b8}).
\par
In Ref.\cite{b1} the conformal transformations were introduced as canonical symmetry transformations  acting on phase manifold. Here we complete the picture by showing that, within the formalism of higher-derivative theories, the conformal transformations can be defined as point transformations acting on configuration space. Using the results of Ref. \cite{b1} we derive the form of point transformations and show that the higher-derivative Lagrangian describing free motion is invariant  (up to a total derivative) under the action of these transformations.
\par
The paper is organized as follows. In Section  \ref{s2} we remind the main results of Ref. \cite{b1} and derive the explicit form of canonical symmetry transformations. In Section \ref{s3} and \ref{s4} the point transformations on configuration space are defined which coincide "on shell" with these introduced in Section \ref{s2} and it is shown that they are Noether symmetries of free higher-derivative Lagrangian with integrals of motion corresponding to the ones on Hamiltonian level (in order to do this we construct an integral of motion for an arbitrary higher-order Lagrangian). Section \ref{s5} is devoted to concise conclusions. Some technicalities are relegated to the Appendix.
\section{Hamiltonian  formalism}
\label{s2}
The $N$-Galilean conformal algebra has the following structure. First, we have the direct sum of $su(2)$ algebra (spanned by $J_k$'s) and $sl(2,\mR)$ one (spanned by $H,D$ and $K$)
\begin{equation}
\label{e1}
\begin{split}
[J^a,J^b]&=i\epsilon_{abc}J^c,\\
[D,H]=iH,\quad [D,K]&=-iK,\quad [K,H]=2iD.
\end{split}
\end{equation}
It is supplemented by $3(N+1)$  (in general case of $d$-dimensional space -- $d(N+1))$ dimensional abelian algebra  which  carries $D^{(1,\frac N 2)}$ representation of $SU(2)\times SL(2,\mR)$ and is spanned by the generators $C_i^a$, $a=1,2,3$; $i=0,1,\ldots,N$. The relevant commutation rules involving $C_i^a$ read
\begin{equation}
\label{e2}
\begin{split}
[J^a,C_j^b]&=i\epsilon_{abc}C^c_j,\quad [H,C_j^a]=-ijC_{j-1}^a,\\
[D,C_j^a]&=i(\frac{N}{2}-j)C_j^a,\quad [K,C_j^a]=i(N-j)C_{j+1}^a.
\end{split}
\end{equation}
For $N$ odd (and also $N$ even in 2+1 dimensions) the algebra defined by eqs. (\ref{e1}) and (\ref{e2}) admits the central extension \cite{b5,b6}.
\begin{equation}
\label{e3}
[C_j^a,C_k^b]=i\delta_{ab}\delta_{N,j+k}(-1)^\frac{k-j+1}{2}j!k!M,
\end{equation}
with $M$ being  additional central generator.
The  above algebra  can be integrated to the group ($SU(2)\times SL(2,\mR))\ltimes \mR^{3N+4}$ where $\mR^{3N+4}$ is nilpotent group and the semidirect product is defined by the $D^{(1,\frac N 2)}\oplus D^{(0,0)}$ representation of $SU(2)\times SL(2,\mR)$.
\par
The question arises what are the dynamical systems exhibiting the symmetry described by $N$-Galilean conformal group. In the case of centrally extended algebra the answer was given in Ref. \cite{b6}. The results obtained there have been generalized  in Ref. \cite{b1} to the space of arbitrary Hamiltonian system  on which our group acts transitively.
Below we discuss the properties of the system constructed by Gomis and Kamimura \cite{b6}; the general case differs by the existence of additional internal degrees of freedom \cite{b1}.
\par
The phase space is parametrized by the canonical variables $q_k^a$, $p_k^a$, $a=1,2,3$, $k=0,1\ldots,\frac {N-1}{2}$, obeying 
\begin{equation}
\label{e4}
\{q_i^a,p_k^b\}=\delta_{ab}\delta_{ik}.
\end{equation}
Defining 
\begin{equation}
\label{e5}
\begin{split}
h(t)&=\frac{1}{2m}(\vec p_{\frac{N-1}{2}})^2+\sum_{k=1}^{\frac{N-1}{2}}\vec q_k\vec p_{k-1},\\
d(t)&=\sum_{k=0}^{\frac{N-1}{2}}(\frac N 2 -k)\vec q_k\vec p_k,\\
k(t)&=\frac{m}{2}\left(\frac{N+1}{2}\right)^2(\vec q_{\frac{N-1}{2}})^2-\sum_{k=0}^{\frac {N-3}{2}}(N-k)(k+1)\vec q_k\vec p_{k+1},\\
\vec j(t)&=\sum_{k=0}^\frac{N-1}{2}\vec q_k\times \vec p_k,
\end{split}
\end{equation}
one finds the Noether charges corresponding  to the generators of Lie algebra (\ref{e1})-(\ref{e3})
\begin{equation}
\label{e6}
\begin{split}
h&=h(t),\\
d&=d(t)-th(t),\\
k&=k(t)-2td(t)+t^2h(t),\\
\vec j&=\vec j(t),\\
\vec c_j&=
\begin{cases}
(-1)^{j-\frac{N-1}{2}}\sum_{k=0}^j\frac{j!}{(j-k)!}t^{j-k}\vec p_k, \quad 0\leq j\leq\frac{N-1}{2},\\
(-1)^{j-\frac{N-1}{2}}\sum_{k=0}^{\frac{N-1}{2}}\frac{j!}{(j-k)!}t^{j-k}\vec p_k+\\
\quad m\sum_{k=\frac{N+1}{2}}^j(-1)^{j-k}\frac{j!}{(j-k)!}t^{j-k}\vec q_{N-k},\quad \frac{N+1}{2}\leq j\leq N.
\end{cases}
\end{split}
\end{equation}
\par
These charges generate the canonical transformations representing the $N$-Galilean conformal group on Hamiltonian level. Computing systematically the infinitesimal action  of all generators we find:
\par
-- for $ \vec c_j:$
\begin{equation}
\label{e7a}
\begin{split}
\delta \vec q_n&=\{\sum_j\vec x_j\vec c_j,\vec q_n\}=\sum_{k=0}^{N-n}(-1)^{k+n-\frac{N+1}{2}}\frac{(k+n)!}{k!}t^k\vec x_{k+n},\\
\delta \vec p_n&=\{\sum_j\vec x_j\vec c_j,\vec p_n\}=m\sum_{k=0}^{n}(-1)^{k}\frac{(k+N-n)!}{k!}t^k\vec x_{k+N-n},
\end{split}
\end{equation}
--  for $h:$
\begin{equation}
\label{e7b}
\begin{split}
\delta \vec q_n&=\{\tau h,\vec q_n\}=-\tau\left(\frac 1 m \delta_{\frac{N-1}{2},n}\vec p_{\frac{N-1}{2}}+(1-\delta_{\frac{N-1}{2},n})\vec q_{n+1}\right),\\
\delta \vec p_n&=\{\tau h,\vec p_n\}= \tau(1-\delta_{n0})\vec p_{n-1},
\end{split}
\end{equation}
--  for $d:$
\begin{equation}
\label{e7c}
\begin{split}
\delta \vec q_n&=\{\lambda d,\vec q_n\}=\lambda\left(-(\frac{N}{2}-n)\vec q_n+t(\frac 1 m \vec p_{\frac{N-1}{2}}\delta_{\frac{N-1}{2},n} +(1-\delta_{\frac{N-1}{2},n})\vec q_{n+1}\right),\\
\delta \vec p_n&=\{\lambda d,\vec p_n\}= \lambda\left(\left (\frac N 2 -n\right ) \vec p_n-t(1-\delta_{n0})\vec p_{n-1}\right),
\end{split}
\end{equation}
--  for $k:$
\begin{equation}
\label{e7d}
\begin{split}
\delta \vec q_n&=\{c k,\vec q_n\}=c\Big( (1-\delta_{n0})n(N-n+1)\vec q_{n-1}+\\
&2t(\frac N 2 -n) \vec q_n-t^2\big( \frac 1 m \delta_{n,\frac{N-1}{2}} \vec p_{\frac{N-1}{2}} +(1- \delta_{n,\frac{N-1}{2}} )\vec q_{n+1} \big) \Big),\\
\delta \vec p_n&=\{ck,\vec p_n\}= c\Big(m\delta_{n\frac{N-1}{2}}\big(\frac{N+1}{2}\big)^2\vec q_{\frac{N-1}{2}}-(1-\delta_{n\frac{N-1}{2}})(N-n)(n+1)\vec p_{n+1}-\\
&2t(\frac N 2 -n)\vec p_n+t^2(1-\delta_{n0})\vec p_{n-1}\Big).\\
\end{split}
\end{equation}
These equations can be integrated out to yield the global transformations.
\section{Lagrangian formalism}
\label{s3}
We want to find the realization of $N$-Galilean conformal group as a group  of symmetry transformations on coordinate space (point transformation). Let us note that the Hamiltonian $h$ given by first  eq. (\ref{e5}) is the Ostrogradski Hamiltonian \cite{b10} corresponding to higher derivative free Lagrangian
\begin{equation}
\label{e8}
L=\frac m 2 \left(\frac{d^{\frac{N+1}{2}}\vec q}{dt^{\frac{N+1}{2}}}\right)^2.
\end{equation}
The procedure of passing from Lagrangian to Hamiltonian formalism is slightly involved \cite{b10}. First, one enlarges the coordinate space  by defining:
\begin{equation}
\label{e9}
\vec q_0=\vec q,\ \vec q_1=\dot{\vec q},\ldots, \vec q_{\frac{N-1}{2}}=\vec q^{(\frac{N-1}{2})};
\end{equation}
then one writes out the Lagrangian 
\begin{equation}
\label{e10}
\tilde L=\frac m 2 \left(\frac{d\vec q_{\frac{N-1}{2}}}{dt}\right)^2+\sum_{k=0}^\frac{N-3}{2}\vec \lambda_k(\dot{\vec q}_k-\vec q _{k+1});
\end{equation}
$\tilde L$ is singular so the Dirac method has to be applied. It appears that all resulting constraints are of the second class which allows to eliminate the Lagrange multipliers $\vec \lambda_k$ and their conjugate momenta. In this way we arrive at the Ostrogradski Hamiltonian.
\par
To find the action of $N$-Galilean conformal group in coordinate space let us remind that the general canonical transformation describes the point transformation  provided the new coordinates are expressible in terms of old ones (with no momenta involved) while the new momenta are the linear functions of old momenta (with coordinate-dependent coefficient); actually they can be more general affine functions if the Lagrangian transforms by a total derivative. However, the above statement is true only provided the time variable remains unchanged. If it changes the momenta can enter the expressions for the variations of coordinate variables provided they appear only in the form of Poisson brackets of Hamiltonian  with coordinates. Then the terms containing momenta can be removed from transformation formulae at the expense of admitting the time variation. The resulting modified  point transformations coincide "on shell" with the initial canonical ones.
\par
A simple inspection of the formulae given in Sec. \ref{s2} shows that the canonical action  of $N$-Galilean conformal group has the above mentioned properties. Therefore, one can define the action of this group on coordinate space which  "on-shell" coincides with the transformations defined in Sec. \ref{e2}. It is not difficult to derive the form of this action. First, let us note that the relations (\ref{e7a})-(\ref{e7d}) yield immediately the global form of transformations generated by $\vec{c}_k$'s:
\begin{equation}
\label{e11}
\begin{split}
t'&=t,\\
\vec {q_n'}(t')&=\vec q_n(t)+\sum_{k=0}^{N-n}(-1)^{k+n-\frac{N+1}{2}}\frac{(k+n)!}{k!}t^k\vec x_{n+k}.
\end{split}
\end{equation}
The global action of $h$ obviously reads 
\begin{equation}
\label{e12}
\begin{split}
t'&=t+\tau,\\
\vec {q_n'}(t')&=\vec q_n(t).
\end{split}
\end{equation}
To find the action of dilatation we rewrite  eq. (\ref{e7c}) in the form 
\begin{equation}
\delta \vec q_n=\lambda (n-\frac N2)\vec q_n+\lambda t\dot{\vec q},
\end{equation}
which can be easy integrated to 
\begin{equation}
\label{e13}
\begin{split}
t'&=e^{-\lambda}t,\\
\vec {q_n'}(t')&=e^{\lambda(n-\frac N2)}\vec q_n(t).
\end{split}
\end{equation}
 The case of conformal transformation is slightly more involved. First, by extracting the coefficient in front of $h(t)$  in the expression defining $k$ we find 
 \begin{equation}
 \label{e14}
 \delta t=ct^2,
 \end{equation}
 which integrates to 
 \begin{equation}
 \label{e15}
t'=\frac{t}{1-ct}\equiv  t(c).
 \end{equation}
 Now, the first eq. (\ref{e7d}) can be written in the form
 \begin{equation}
 \label{e16}
 \delta\vec{q}_n=c\left(n(N-n-1)\vec q_{n-1}+2t(\frac N2 -n)\vec q_n -t^2\dot{\vec q}_n\right).
 \end{equation}
The last term on the right-hand side is responsible for time variation. One can get rid of this term by replacing the time variable by its "running" value (\ref{e15}). In this way we arrive at the following equations
 \begin{equation}
 \label{e17}
\frac{d\vec q_n(t(c),c)}{dc}=n(N-n+1)\vec{q}_{n-1}(t(c),c)+\frac{2t}{1-ct}(\frac N2-n)\vec q_n(t(c),c).
 \end{equation}
 It is not difficult to integrate eq. (\ref{e17}). The result reads
  \begin{equation}
 \label{e18}
 \vec{q_n'}(t')=\sum_{k=0}^n\dbinom{n}{k}\frac{(N+k-1)!}{(N-1)!}\frac{c^k}{(1-ct)^{N+k}}\vec q_{n-k}(t).
 \end{equation}
Finally, the action of rotation subgroup is standard.
\par
Let us note that in all cases the following important property holds: 
\begin{equation}
\label{e19}
\vec q_{n+1}=\dot{\vec q}_n \textrm{\quad  implies \quad }{\vec {q'}_{n+1}}(t')=\frac{d \vec {q_n'}(t')}{dt'}.
\end{equation}
It allows us to reduce the action of the group under  consideration to that on the variables $t$ and $\vec q=\vec q_0$. One finds
\begin{equation}
\label{e20}
\begin{split}
t'=t, \quad
\vec {q'}(t')=\vec q(t)+\sum_{k=0}^{N}(-1)^{k-\frac{N+1}{2}}t^k\vec x_k;
\end{split}
\end{equation}
\begin{equation}
\label{e21}
\begin{split}
t'=t+\tau,\quad
\vec {q'}(t')=\vec q(t);
\end{split}
\end{equation}
\begin{equation}
\label{e22}
\begin{split}
t'=e^{-\lambda}t,\quad
\vec {q'}(t')=e^{-\lambda\frac N2}\vec q(t);
\end{split}
\end{equation}
\begin{equation}
\label{e23}
\begin{split}
t'=\frac{t}{1-ct},\quad
\vec {q'}(t')=\frac{\vec q(t)}{(1-ct)^N};
\end{split}
\end{equation}
as the counterparts  of eqs. (\ref{e11}) (\ref{e12}) (\ref{e13}) (\ref{e15}) (\ref{e18}), respectively.
\par
It is shown in Appendix that in all cases $Ldt$  (with $L$ given by eq. (\ref{e12})) is invariant, up to an exact differential, under all the above transformations.
\par
Eqs. (\ref{e20})-(\ref{e23}) allow us to write out the differential realization of the algebra (\ref{e1}), (\ref{e2}). It reads
\begin{equation}
\label{e24}
\begin{split}
\hat H=i\frac{\partial }{\partial t}, \quad \hat D=-i\left(\frac N2\vec q\frac{\partial}{\partial \vec q}+t\frac{\partial}{\partial t}\right),\\
\quad \hat K= i\left(Nt\vec q\frac{\partial }{\partial q}+t^2\frac{\partial}{\partial t} \right),\ 
\hat{\vec C}_k=i(-1)^{k-\frac{N-1}{2}}t^k\frac{\partial }{\partial \vec q}.  
\end{split}
\end{equation} 
Let us also note that $\hat{\vec C}_k$ commutate with each other. This is  due to the fact that $M$, being central element, act trivially in coadjoint representation.

\par
In order to complete the picture we will find  all integrals of motion corresponding to the transformations (\ref{e20})-(\ref{e23}) and  compare them with the ones defined on the Hamiltonian level (\ref{e6}). First, let us note that for  infinitesimal   transformations 
\begin{equation}
\label{ee1}
\vec {q'}=\vec q+\epsilon\vec\chi(q,t),\quad t'=t+\epsilon g(t),
\end{equation}
and an arbitrary higher-order Lagrangian $L=L(\vec q,\dot{\vec q},\ldots,{\vec {q}}^{(R)})$, $R>1$, the symmetry condition 
\begin{equation}
\label{ee2}
L(\vec {q'}(t'),\frac{d{\vec {q'}}}{dt'},\ldots,\frac{d^R\vec {q'}}{dt'^R})\frac{dt'}{dt}= L(\vec {q}(t),\frac{d{\vec {q}}}{dt},\ldots,\frac{d^R\vec {q}}{dt^R})+\epsilon \frac{d}{dt}(\delta f),
\end{equation}
implies 
\begin{equation}
\label{ee3}
\epsilon \frac{d}{dt}(\delta f)- \sum_{n=0}^R\frac{\partial L}{\partial \vec q ^{(n)}}\delta(\frac{d^n \vec q}{dt^n})-\epsilon\dot g L=0.
\end{equation}
For $n>0$ the following identity holds 
\begin{equation}
\label{ee4}
\delta(\frac{d^n \vec q}{dt^n})=-\epsilon \sum_{k=0}^{n-1}\frac{d ^k}{dt^k}(\dot g\vec q^{(n-k)})+\epsilon \vec\chi^{(n)},
\end{equation}
while for $n>1$, $k>0$ we get
\begin{equation}
\label{ee5}
\begin{split}
\frac{d^k}{dt^k}(\dot g \vec q^{(n-k)})\frac{\partial L}{\partial \vec q ^{(n)}}&=\frac{d}{dt}\left(\sum_{l=0}^{k-1}(-1)^l\frac{d^{k-l-1}}{dt^{k-l-1}}(\dot g \vec q^{(n-k)})\left(\frac{d^l}{dt ^l}\right)\left(\frac{\partial L}{\partial \vec q^{(n)}}\right)\right)+\\
&(-1)^k\dot g \vec q^{(n-k)}\frac{d^k}{dt^k}\left(\frac{\partial L}{\partial \vec q ^{(n)}}\right).
\end{split}
\end{equation}
Using eqs. (\ref{ee4}) and (\ref{ee5}) one finds 
\begin{equation}
\label{ee6}
\begin{split}
&\sum_{n=1}^R\frac{\partial L}{\partial \vec q ^{(n)}}\delta(\frac{d^n \vec q}{dt^n})=-\epsilon\dot g\sum_{n=1}^R\sum_{k=0}^{n-1}(-1)^k\frac{d^k}{dt^k}\left(\frac{\partial L}{\partial \vec q ^{(n)}}\right) \vec q ^{(n-k)}+\epsilon\sum_{n=1}^R\frac{\partial L}{\partial \vec q^{(n)}}\vec \chi^{(n)}\\
&-\epsilon\sum_{n=2}^R\sum_{k=1}^{n-1}\frac{d}{dt}\left(\sum_{l=0}^{k-1}(-1)^l\frac{d^{k-l-1}}{dt^{k-l-1}}(\dot g \vec q^{(n-k)})\left(\frac{d^l}{dt ^l}\right)\frac{\partial L}{\partial \vec q^{(n)}}\right).
\end{split}
\end{equation}
On the other hand the Ostrogradski Hamiltonian for $L$  can be written in the form
\begin{equation}
\label{ee7}
H=\sum_{l=0}^{R-1}\vec p_l\vec q^{(l+1)}-L,
\end{equation}
where 
\begin{equation}
\label{ee8}
\vec p_n=\sum_{j=0}^{R-n-1}\left(-\frac{d}{dt}\right)^j\left(\frac{\partial L}{\partial \vec q^{(n+j+1)}}\right),\quad n=0,1,\ldots,R-1;
\end{equation}
consequently the first term on the r.h.s. of eq. (\ref{ee6}) can be rewritten as follows 
\begin{equation}
\label{ee9}
\begin{split}
&-\epsilon\dot g\sum_{n=1}^R\sum_{k=0}^{n-1}(-1)^k\frac{d^k}{dt^k}\left(\frac{\partial L}{\partial \vec q ^{(n)}}\right) \vec q ^{(n-k)}=-\epsilon\dot g\sum_{l=1}^R\sum_{j=0}^{R-l}\left(-\frac{d}{dt}\right)^j\left(\frac{\partial L}{\partial \vec q^{(l+j)}}\right)\vec q^{(l)}=\\
&-\epsilon\dot g\sum_{l=0}^{R-1}\sum_{j=0}^{R-l-1}\left(-\frac{d}{dt}\right)^j\left(\frac{\partial L}{\partial \vec q^{(l+j+1)}}\right)\vec q^{(l+1)}=-\epsilon\frac{d}{dt}{( gH)}- \epsilon \dot gL,
\end{split}
\end{equation}
where $H$ is expressed in terms of $\vec q$ and their time derivatives. 
Substituting this result into eq. (\ref{ee6}) and  using eq. (\ref{ee3}) we obtain the following  equation
\begin{equation}
\label{ee10}
\begin{split}
&\epsilon \frac{d}{dt}\left(\delta f+H\dot g+\sum_{n=2}^R\sum_{k=1}^{n-1}\sum_{l=0}^{k-1}(-1)^l\frac{d^{k-l-1}}{dt^{k-l-1}}(\dot g \vec q^{(n-k)})\left(\frac{d^l}{dt ^l}\right)\frac{\partial L}{\partial \vec q^{(n)}}\right)-\\
&\epsilon \sum_{n=0}^R\frac{\partial L}{\partial \vec q^{(n)}}\vec \chi^{(n)}=0.
\end{split}
\end{equation}
Moreover, one checks that
\begin{equation}
\label{ee11}
\sum_{n=0}^R\frac{\partial L}{\partial \vec q^{(n)}}\vec \chi^{(n)}=\frac{d}{dt}\left(\sum_{k=0}^{R-1}\vec p_k\vec\chi ^{(k)}\right)+\vec \chi\sum_{k=0}^R\left(-\frac{d}{dt}\right)^k\left(\frac{\partial L}{\partial \vec q^{(k)}}\right).
\end{equation}
Together with eq. (\ref{ee10}) this leads to the following   integral of motion  
\begin{equation}
\label{ee12}
C=Hg-\sum_{k=0}^{R-1}\vec p_k\vec\chi ^{(k)}+\sum_{n=2}^R\sum_{k=1}^{n-1}\sum_{l=0}^{k-1}\frac{d^{k-l-1}}{dt^{k-l-1}}(\dot g \vec q^{(n-k)})\left(-\frac{d^l}{dt ^l}\right)\left(\frac{\partial L}{\partial \vec q^{(n)}}\right) +\delta f.
\end{equation}
\par 
Now, let us apply these general formulae  to our Lagrangian (\ref{e8}) and symmetry transformations (\ref{e20})-(\ref{e23}).
In this case the generalized  momenta (\ref{ee8}) and  the Hamiltonian $H$, when written in terms of $\vec q$'s, read
\begin{equation}
\label{ee13}
\vec p_n=m(-1)^{\frac{N-1}{2}-n}\vec q^{(N-n)},\quad  n=0,1,\ldots,\frac{N-1}{2},
\end{equation}
\begin{equation}
\label{ee14}
H=\sum_{n=0}^{\frac{N-3}{2}}(-1)^{\frac{N-1}{2}-n}\vec q^{(N-n)}\vec q ^{(n+1)}+\frac m2 (\vec q ^{(\frac{N+1}{2})})^2.
\end{equation}
Let us now find the  integrals of motion.
For the transformations  (\ref{e20}),  $g_k=0$ and $\vec \chi_k=(-1)^{k-\frac{N+1}{2}}t^k$, $k=0,\ldots,N$ while the functions $ \delta\vec f_k$ are of the form 
(see Appendix, eq. (\ref{e26}) for small $\vec x$)
\begin{equation}
\label{ee15}
\begin{split}
&\delta\vec f_k=0,\ k=0,\ldots,\frac{N-1}{2};\\
&\delta\vec f_k=m\sum_{n=\frac{N+1}{2}}^k(-t)^{k-n}\frac{k!}{(k-n)!}\vec q^{(N-n)},\ k=\frac{N+1}{2},\ldots,N.
\end{split}
\end{equation}
Due to eq. (\ref{ee12}) the corresponding integrals of motion are of the form 
\begin{equation}
\label{ee16}
\vec C_k=m\sum_{n=0}^k(-t)^{k-n}\frac{k!}{(k-n)!} \vec q ^{(N-k)},\ k=0,\ldots,N.
\end{equation}  
For time translations obviously  $C=H$. 
\par
Similarly, for dilatations (\ref{e22}), $\delta f=0$. Substituting  $g=-t$, $\vec \chi=\frac{-N}{2}\vec q$   into eq.
(\ref{ee12}) we obtain the following integral of motion 
\begin{equation}
D=-tH+D(t)=-tH+m\sum_{k=0}^{\frac{N-1}{2}}(-1)^{\frac{N-1}{2}-k}(\frac N2 -k)\vec q^{(N-k)}\vec q^{(k)}.
\end{equation}
For conformal transformations (\ref{e23}), eq. (\ref{e28}) implies $\delta f=\frac m2\left(\frac{N+1}{2}\right)^2\left(\vec q^{(\frac{N-1}{2})}\right)^2$. Moreover  $g=t^2$ and $\vec \chi=tN\vec q$. Using this we find, after some computations, the corresponding integral of motion 
\begin{equation}
\label{ee17}
\begin{split}
K&=t^2H-2tD(t)+K(t)=t^2H-2tD(t)+\\
&m\sum_{j=0}^{\frac{N-3}{2}}(j+1)(N-j)(-1)^{\frac{N-1}{2}-j}\vec q^{(j)}\vec q^{(N-j+1)}+ 
\frac m2\left(\frac{N+1}{2}\right)^2\left(\vec q^{(\frac{N-1}{2})}\right)^2.
\end{split}
\end{equation}
Finally, by considering rotations one gets the following expression for angular momentum 
\begin{equation}
\label{ee18}
\vec J=m\sum_{k=0}^{\frac{N-1}{2}}(-1)^{\frac{N-1}{2}-k}\vec q^{(k)}\times \vec q ^{(N-k)}.
\end{equation}
Concluding, let us note that, as in the case of first order theory, all  integrals of motion (\ref{ee16})-(\ref{ee18})  can be obtained from the ones on the Hamiltonian level (see, eqs. (\ref{e6})) by expressing Ostrogradski momenta (\ref{ee13}) in terms of $\vec q$ and it time derivatives. 
\section{Two-dimensional case}
\label{s4}
As we have mentioned, in the case of  dimension $2$ and  $N$ even, there also exists the central extension of $N$-GCA (\ref{e1}) and (\ref{e2}). It is given by
\begin{equation}
\label{ee19}
[C_j^a,C_k^b]=-i\epsilon^{ab}\delta^{N,j+k}(-1)^{\frac{j-k}{2}}k!j!M,
\end{equation}
where $\quad a,b=1,2,\quad j,k=0,1,\ldots,N$. 
Neglecting "internal" degrees of freedom general  Hamiltonian  systems with this symmetry is of the form (see, \cite{b1})
\begin{equation}
\label{ee20}
\begin{split}
\{q_j^a,p_k^b\}&=\delta^{ab}\delta_{jk},\quad j,k=0,\ldots ,\frac N2-1,\quad a,b=1,2;\\
\{q_{\frac N2}^a,q^b_{\frac N2}\}&=\frac 1m\epsilon^{ba},\quad a,b=1,2.
\end{split}
\end{equation}
Denoting  $p_{\frac N2}^a=\frac m2\epsilon ^{ba}q_{\frac N2}^b$  and
\begin{equation}
\label{ee21}
\begin{split}
h(t)&=\sum_{k=0}^{\frac{N}{2}-1}\vec p_k\vec q_{k+1},\\
d(t)&=\sum_{k=0}^{\frac{N}{2}-1}(\frac N2 -k)\vec p_k \vec q_k,\\
k(t)&=-\sum_{k=1}^{\frac{N}{2}-1}(N-k+1)k\vec p_k\vec q_{k-1}-N(\frac N2+1)\vec q_{\frac N2 -1}\vec p_{\frac N2},\\
j(t)&=\sum_{k=0}^{\frac{N}{2}}\vec q_k\times \vec p_k,
\end{split}
\end{equation}
one can find Noether's charges:
\begin{equation}
\label{ee22}
\begin{split}
h&=h(t),\\
d&=d(t)-th(t),\\
k&=k(t)-2td(t)+t^2h(t),\\
j&=j(t),\\
c_j^a&=
\begin{cases}
(-1)^{j+\frac{N}{2}+1}\sum_{k=0}^j\frac{j!}{(j-k)!}t^{j-k} p_k^a, \quad 0\leq j\leq\frac{N}{2}-1\\
(-1)^{j+\frac{N}{2}+1}\sum_{k=0}^{\frac{N}{2}-1}\frac{j!}{(j-k)!}t^{j-k} p_k^a+\\
\quad m\sum_{k=\frac{N}{2}}^j(-1)^{j-k}\frac{j!}{(j-k)!}t^{j-k}\epsilon_{ab} q_{N-k}^b,\quad \frac{N}{2}\leq j\leq N.
\end{cases}
\end{split}
\end{equation}
Of course, we can repeat the preceding  considerations. 
For example, infinitesimal  transformations of $\vec q_0$ are of the form:
\begin{subequations}
\label{ee23}
\par 
--  for $ \vec c_j:$
\begin{equation}
\begin{split}
\delta \vec q_0&=\{\sum_j\vec x_j\vec c_j,\vec q_0\}=\sum_{j=0}^Nt^j(-1)^{j-\frac N2}\vec x_j,\\
\end{split}
\end{equation}
--  for $h:$
\begin{equation}
\begin{split}
\delta \vec q_0&=\{\tau h,\vec q_0\}=-\tau\dot \vec q_1,
\end{split}
\end{equation}
--  for $d:$
\begin{equation}
\begin{split}
\delta \vec q_0&=\{\lambda d,\vec q_0\}=\lambda(t \vec q_1-\frac N2\vec q_0),\\
\end{split}
\end{equation}
--  for $k:$
\begin{equation}
\begin{split}
\delta \vec q_0&=\{c k,\vec q_0\}=c(t\vec q_0N-t^2\vec q_1).
\end{split}
\end{equation}
\end{subequations}
\par
In terms of the variable $\vec q=\vec q_0$ the Lagrangian corresponding to the Ostrogradski Hamiltonian $h$  reads  
\begin{equation}
\label{ee24}
L=\frac m 2\epsilon_{ab}  \frac{d^{\frac{N-1}{2}} q^a}{dt^{\frac{N-1}{2}}}\frac{d^{\frac{N+1}{2}} q^b}{dt^{\frac{N+1}{2}}}.
\end{equation}
As previously, one can show that the action of $N$-GCA can be translated into coordinate space $\vec q$  and the symmetry transformations, corresponding to (\ref{ee23}), are of the form (cf. eqs. (\ref{e20})-(\ref{e23}) for the case of $N$ odd)
\begin{subequations}
\begin{equation}
t'=t, \quad
\vec {q'}(t')=\vec q(t)+\sum_{k=0}^{N}(-1)^{k-\frac{N}{2}}t^k\vec x_k;
\end{equation}
\begin{equation}
t'=t+\tau,\quad
\vec {q'}(t')=\vec q(t);
\end{equation}
\begin{equation}
t'=e^{-\lambda}t,\quad
\vec {q'}(t')=e^{-\lambda\frac N2}\vec q(t);
\end{equation}
\begin{equation}
t'=\frac{t}{1-ct},\quad
\vec {q'}(t')=\frac{\vec q(t)}{(1-ct)^N}.
\end{equation}
\end{subequations}
Differential realization of the algebra is the same as in the case of $N$ odd  (there is  not central extension  on Lagrangian level).
\section{Conclusions}
\label{s5}
In this paper we completed the picture by providing dynamical realization of $N$-Galilean conformal symmetries, in the case when the relevant algebra admits central extension. It has been shown previously \cite{b1,b6}, that the generic Hamiltonian system on which the $N$-Galilean conformal group acts transitively as a group of canonical transformation is described by Ostrogradski Hamiltonian  (modulo the dynamics of "internal"  degrees of freedom).  Here we defined the point transformations acting on coordinate  space  which coincide "on-shell" with the above canonical ones. They provide the Noether symmetries of free Lagrangian containing  $\frac{N+1}{2}$-th order time derivatives as well as the Noether's charges coinciding with those obtained on Hamiltonian level. 
It can be further  shown \cite{b11} that $N$-Galilean conformal group is the maximal group of  Noether symmetries of free  higher-derivative Lagrangian. This generalizes the Niederer's result \cite{b9} concerning the Schr\"odinger group, $N=1$ (actually, Niederer proved it on quantum level)
\par
{\bf Acknowledgments}
The authors would like to thank  Cezary Gonera, Piotr Kosi\'nski and Pawe\l\  Ma\'slanka for helpful discussions and  useful remarks. This work  is supported  in part by  MNiSzW grant No. N202331139.
\section{Appendix}
Let us now show explicitly  that under the transformations of $N$-Galilean conformal group the Lagrangian (\ref{e8})  transforms by total derivative; more precisely we prove that
\begin{equation}
\label{e25}
\frac m2\left(\frac{d^{\frac{N+1}{2}}\vec{q'}}{dt'^{\frac{N+1}{2}}}\right)^2\frac{dt'}{dt}=\frac m2\left(\frac{d^{\frac{N+1}{2}}\vec{q}}{dt^{\frac{N+1}{2}}}\right)^2+\frac{d f}{dt},
\end{equation}
where $f$ is a function of $\vec q$ and its time derivatives up to $\frac{N-1}{2}$ order.
 Using the formulae of Sec. {\ref{s2} we see that $f=0$ for time translations, dilatations and  rotations. For transformations (\ref{e20}) the  functions $ f_k$ are of the form
 \begin{equation}
 \label{e26}
 \begin{split}
 & f_k=0\quad k=0,\ldots,\frac{N-1}{2} \\
 & f_k=m\sum_{n=\frac{N+1}{2}}^k(-t)^{k-n}\frac{k!}{(k-n)!}\vec q^{(N-n)}\vec x+\\
&\frac m2 \left(\frac{k!}{(k-\frac{N+1}{2})!}\right)^2\frac{1}{2k-N}t^{2k-N}\vec x^2,\quad k=\frac{N+1}{2},\ldots , N.
 \end{split}
 \end{equation}
 For the  conformal transformation (\ref{e23}) the condition  (\ref{e25})  takes the form 
 \begin{equation}
 \label{e27}
 \begin{split}
 &\frac m2\left(\frac{d^{\frac{N+1}{2}}\vec{q'}}{dt'^{\frac{N+1}{2}}}\right)^2\frac{dt'}{dt}-\frac m2\left(\frac{d^{\frac{N+1}{2}}\vec{q}}{dt^{\frac{N+1}{2}}}\right)^2=\\
&\sum_{\overset{l,l'=0}{l+l'<N+1}}^{\frac{N+1}{2}}\frac{(\frac{N+1}{2})^2(N-l)!(N-l')!c^{N+1-l-l'}}{l!l'!(\frac{N+1}{2}-l)(\frac{N+1}{2}-l')!(1-ct)^{N+1-l-l'}}\vec q^{(l)}\vec q^{(l')}.
 \end{split}
 \end{equation}
 Let us put 
 \begin{equation}
 \label{e28}
 f=\frac m2\left(\frac{N+1}{2}\right)^2\sum_{l,l'=0}^{\frac{N-1}{2}}\frac{a(l,l')c^{N-l-l'}}{(1-ct)^{N-l-l'}}\vec q^{(l)}\vec q^{(l')}.
 \end{equation}
 Then eqs. (\ref{e27}) and (\ref{e28}) imply 
 \begin{equation}
 \label{e29}
 \begin{split}
(N-l-l')a(l,l')+a(l-1,l')+a(l,l'-1)=
\frac{(N-l)!(N-l')!}{l!l'!(\frac{N+1}{2}-l)(\frac{N+1}{2}-l')!}.
\end{split}
 \end{equation}
Eq. (\ref{e25}) is established once we show that this recurrence has the solution. To this end we put 
 \begin{equation}
 \label{e30}
 a(l,l')=(N-l-l'-1)!d(l,l').
 \end{equation}
In terms of  $d(l,l')$ eq.  (\ref{e29})  reads
 \begin{equation}
 \label{e31}
d(l,l')+d(l-1,l')+d(l,l'-1)=
\frac{1}{N!}C_{ll'}^N\prod_{k=\frac{N+3}{2}}^N(k-l)\prod_{k'=\frac{N+3}{2}}^N(k'-l'),
 \end{equation}
where
 \begin{equation}
 \label{e32}
 C_{l_1l_2}^n=\frac{n!}{l_1!l_2!(n-l_1-l_2)!}.
 \end{equation}
Note that the following  identity holds
 \begin{equation}
 \label{e33}
 C_{l_1l_2}^n+C_{(l_1-1)l_2}^n+C_{l_1(l_2-1)}^n=C_{l_1l_2}^{n+1}.
 \end{equation}
The product $\prod_{k=\frac{N+3}{2}}^N(k-l)$ is a polynomial in $l$ of degree $\frac{N-1}{2}$ with $N$-dependent coefficients. It is easy to show that it can be rewritten as
 \begin{equation}
 \label{e34}
\prod_{k=\frac{N+3}{2}}^N(k-l)=\sum_{n=0}^{\frac{N+1}{2}}\beta_n(N)\prod_{k=0}^{n-1}(l-k),
 \end{equation}
 for some $\beta_n(N)$'s. Using (\ref{e34}) one can rewrite eq. (\ref{e31}) as
 \begin{equation}
 \label{e35}
d(l,l')+d(l-1,l')+d(l,l'-1)=
\sum_{n,n'=0}^{\frac{N-1}{2}}\gamma_{n,n'}(N)C^{N-n-n'}_{l-n \ l'-n'}.
 \end{equation}
By virtue of the identity (\ref{e33}) we find 
 \begin{equation}
 \label{e36}
d(l,l')=\sum_{n,n'=0}^{\frac{N-1}{2}}\gamma_{n,n'}(N)C^{N-n-n'-1}_{l-n \ l'-n'},
 \end{equation}
so eq.(\ref{e25}) holds.

\end{document}